\documentclass[twocolumn]{elsart}
\usepackage{natbib}
\begin{document}
%
%
\begin{frontmatter}
\title{The challenge of hybridization}

\author[mca]{Massimo Caccia}

\address[mca]{Universita' dell'Insubria, Via Valleggio 11, Como, Italy}

\begin{abstract}
Hybridization of pixel detector systems has to satisfy tight requirements: high
yield, long term reliability, mechanical stability, thermal compliance and
robustness have to go together with low passive mass added to the system,
radiation hardness, flexibility in the technology end eventually low cost. The
current technologies for the interconnection of the front-end chips and the
sensor are reviewed and compared, together with the solutions for the
interface to the far-end electronics.
\end{abstract}

\begin{keyword}
Hybridization; bump bonding; Silicon detector; pixel
\end{keyword}
\end{frontmatter}

\section{Introduction}

Moving from a detector prototype to a full scale system, integrated in a large
experimental facility, requires a full production chain optimization where
hybridization plays a crucial role.
As a case study, the pixel detector for the DELPHI experiment at LEP may be
considered \cite{pierre}. As an upgrade of the existing Silicon strip vertex detector, pixel
sensor endcaps were installed, covering the low polar angle region (${\rm
10-25^{o}}$) and improving the hermeticity \cite{lep2}. The detector was installed in 1996 and 
it has been running for $\approx$ 8 months a year since 1997 without hardware
interventions or dismantling, achieving the design performances.
The DELPHI pixel detector represents a succesful pioneering project;
nevertheless, it appears as being simple compared to the next generation of
detectors planned for the future experiments at the Large Hadron Collider at
CERN and the Tevatron at Fermilab \cite{atlas_ge,cms_ge,alice_ge,btev_ge}:

\begin{itemize}

\item the DELPHI detector is made up of 152 modules, for ${\rm \approx 1.2 \times
10^{6} }$ pixels. In the future detectors, the number of modules increases by a
factor 20 and the number of pixels by two orders of magnitude.

\item DELPHI pixels were meant for tracking
in a low redundancy region rather than vertexing. This is the reason for
the pixel cell size to be ${\rm 330 \times 330 \mu m^{2}}$ in DELPHI, 
scaled down to a ${\rm 50 \mu m }$ minimal pitch and area ${\rm \approx 10^{-4}
cm^{2}}$ in the future applications where vertexing is a must. For the same
region, the material budget at LHC and Tevatron is quite tight. At the moment, 
the sensor and front end chip assembly, the substrate for the power bus and
control lines and the mechanical support contribute to the passive material
(in radiation length) according to the ratio ${\rm 1:0.4:1.5}$. If the mechanics
is still dominating, the bus substrate is far from being negligible.

\item LEP is a radiation soft environment and levels below 1 Krad/year are
measured. At LHC, the dose is expected to increase by three orders of
magnitude,
for an equivalent flux of 1 Mev energy neutrons 
${\rm \approx 10^{14} cm^{-2}}$ per year at the vertex
detector radii. As a consequence,
all of the hybridization and assembly technologies have to be qualified as
radiation hard. Moreover, while the DELPHI pixels are operated at room
temperature, the future detectors will run at around 
${\rm -10~^{o}C}$ and effects related to mismatches among the coefficients of
thermal expansion  of the assembly are critical.

\item the DELPHI pixels are clocked at ${\rm 5 MHz}$ 
while the future experiments requires a ${\rm 40 MHz}$ clock, so that the
properties of the stripline transmission lines have to be qualified. 
\end{itemize}

 
Irrespective of
the rather soft boundary conditions outlined above, the final production yield
of the DELPHI pixels was at the ${\rm 40\%}$ level \cite{johann}.
In particular, 
bump bonding and assembly
were  rather critical, with efficiencies at the ${\rm 80\%~and~55\%}$ 
respectively.
The future pixel detectors have a ${\rm 70\%}$ production yield benchmark
and much more critical constraints; 
achieving it
without degrading the expected performances and fitting the time scale is the
real challenge of hybridization.

In the following, bump bonding issues and techniques for the signal and power
line routings are reviewed, outlining the main advantages and disadvantages.

\section{Front-end chip and sensor interconnection}
Due to the contact density \\
(${\rm 5000 - 10000 cm^{-2}}$), minimum pitch (${\rm 50 \mu m}$),
and the required single bump failure rate (${\rm <10^{-4}}$) the only interconnection technique
is bump bonding of flipped chips. Stencil printing of conductive glue has been proven to work
at larger pitch \cite{pindo}; anisotropic conductive film is also used by industry at larger pitches,
especially for LCD packaging; stud bonding may be considered for prototyping as single dies can
be bonded but it is not an option by now for a batch production. 

The choice of the metal bump, whether it is 
going to be Indium or a solder paste, determines the process characteristics in terms of the 
under bump metallization (UBM), the bonding mechanism and parameters, 
the mechanical properties and the bump aspect ratio. 

\subsection{Solder bumps}
Solder bumps have been introduced about 30 years ago by IBM to overcome the limits of wire
bonding. The steps of the original IBM C4 process (Controlled Collapse Chip Connection \cite{ibm})
may be summarized as follows:
\begin{itemize}
\item after the wafer is cleaned, the under bump metal layers are grown on the final pad by
evaporation through a molybdenum mask, with ultimate via size of ${\rm 50 \mu m}$.
The UBM is meant to provide:
\begin{itemize}
\item an adhesion layer ("active metal", usually Cr or Ti:W) on top of the final metal pad
\item a barrier layer (phased Cr-Cu) to prevent the solder dissolving the active metal
\item a solder wettable layer (Cu or Ni)
\item an oxide prevention layer (Au)
\end{itemize}
\item a lead rich solder (either 97Pb/Sn or 95Pb/Sn) is evaporated through the mask, for 
bumps diameter and height in the ${\rm 100-125 \mu m}$ range
\item the wafer is heated up to the phase transition temperature,
that depends on the solder composition and may achieve a maximum of ${\rm \approx 350^{o}C}$ 
("reflow").
Because of surface tension, the bump will undergo a predictable collapse to an equilibrium
shape corresponding to a truncated sphere \cite{coll}. Reflow may occurr in presence of a suitable flux or
in an organic compound bath.
\end{itemize}

The original IBM prescription presents three major limitations \cite{multi,flipchip,ele0}:
because of the mismatch between the Silicon and Molybdenum coefficients of thermal
expansion the process cannot be effectively scaled up past 5 inch wafers;
the finite mask thickness and the mask alignment procedure limits the typical pitch
to ${\rm 250 \mu m}$;
eutectic Pb/Sn solder
cannot be evaporated due to the low vapour pressure of the Sn. The use of eutectic solder bumps (37Pb/Sn, with a reflow temperature of ${230^{o}C}$)) 
is pursued by industry as it provides the option of mounting the IC onto the circuit board
using the same techniques employed for the surface mountable components. 
These limitations are currently overcome by electroplating the UBM and the solder in a pattern
defined by photoresist masks \cite{ele0,electro}. 
Critical parameters in this case are solder height and alloy
uniformity, linked to the control of the plating current and solution flow; solder voiding has
also been reported \cite{flipchip,ele0},
connected to hydrogen entrapment in the plating solution. Industrial
standards correspond to ${\rm 150 \mu m}$ pitch for ${\rm 70-80 \mu m}$ footprint.

Surface tension effects during the reflow are essential during the flip-chip attachment, as
they induce self-alignment and planarization of the chip with respect to the substrate \cite{self0}.
As a consequence, effective bonding is achieved for overlaps between the bond and the 
wettable substrate pad exceeding ${\rm 25\%}$ of the pad area.

The main issues making the solder bump techniques appealing for the interconnection of sensors
and front-end chips are: bump uniformity and self-alignment properties, resulting in a low
single bump failure rate; the optimal height/pitch ratio, e.g. ${\rm 15-20 \mu m}$ height for a
${\rm 50 \mu m}$ pitch \cite{gec1,izm1}, limiting the parasitic capacitance also in a single sided process; the
excellent electrical properties (contact resistance at the ${\rm m \Omega}$ level \cite{izm1}).
On the other hand, there are critical issues that should not be underestimated as they could
affect the interconnection reliability: the UBM is quite complex and a not perfect control of
the metallurgy can result in a poor adhesion; the process requires relatively high temperature
steps, that might not be tolerated by radiation hard processes; the reflow chemistry should
guarantee wire bondability of the I/O pads. Thermal fatigue and high stress induced by
mismatches in the coefficient of thermal expansions of the assembly are also of concern
\cite{crack0,crack1}; the
mechanical properties (in terms of ultimate tensile and shear stress) depends on the bump
dimensions, on the UBM characteristics and the substrate material but in general fatigue life is
increased by the use of a filler between the substrate and the chip \cite{flipchip}, possibly not tolerated
in assemblying detectors. In current high energy physics applications,
up to 16 chips are connected to same substrate;  replacing a
faulty chip ("reworking") could considerably improve the yield. For instance, in the DELPHI pixel
detector the module production yield after bump bonding of 16 known good chips was ${\rm
80\%}$ and a good part of the faults were on a single chip.
Reworking chips as close as ${\rm 50 \mu m}$ is not trivial in term of the
required temperature and stress and mostly because residual solder has to be 
cleaned before the new placement occurs.   

Nevertheless, 
excellent results were obtained on small volume detector assemblies
for the WA97 \cite{eric} and DELPHI experiments \cite{lep2} 
and on prototypes for the ATLAS experiment \cite{wolf}, with single
bump failure rate in the ${\rm 10^{-4}-10^{-5}}$ range and pitch down to ${\rm 50 \mu m}$. The
results obtained so far proves the 
technology is mature and solutions to specific problems certainly benefits of the enourmous
knowledge gained to define an industrial standard.
  
\subsection{Indium bumps}
Indium bumping technology has been developed for Infrared Sensor assemblies
as In retains its mechanical properties also at
liquid Nitrogen temperature where the sensors are operated. In bumps are in general
grown by evaporation through a patterned photoresist, after a proper UBM. The UBM may be 
simple and limited to a single Cr adhesion layer and bump pitches in the ${\rm 20-30 \mu m}$ are
standard \cite{fiorello}. 
Chemical etching of the photoresist and evaporation are critical parameters in the process,
as a bad control may result in Indium attachment to the via walls, making the lift-off 
not effective \cite{multi,fiorello}. 
As a result of it, In bumps have a small height/pitch ratio (e.g. ${\rm 
7 \mu m/50 \mu m}$) and bumping on both sides of the dies being assembled may be necessary
\cite{gemme}.
Indium belongs to the third group of the periodic table and it features 
a remarkable tendency to form an oxide crust (its electrode potential is +0.38 V, to be compared to
+1.66 V for Al).
As a result of it, bump reflow is extremely difficult even if possible
according to proprietary processes (see for instance \cite{roland}).
Solid state diffusion bonding occurs at 
moderate pressure (${\rm 10^{-2}N/bump}$) and temperature (${\rm 20-100^{o}C}$)
\cite{fiorello,gemme}, but it requires
excellent planarity (at the ${\rm 0.1 mrad}$ level) and bump uniformity.
The latter is the most critical point of the In bumping technology, together with a possible
electronics noise increase due to the "small" bumps.
The remarkable points may be summarized in the simple UBM, the low bonding
temperature and the fact that industry standards match the requirements for detector assembly. 
Reworking appears simpler as the In left
over on the substrate may be used for the fresh substrate attachment.
Indium
plastic properties should help in system assembly but this has to be traded off with a lower
shear and tensile strength with respect to solder bumps.
The ultimate tensile strength for SnPb bumps has been measured to be in the 
${\rm 25-50 MPa}$ range \cite{crack0}, to be
compared to ${\rm 1.9 MPa}$ for In at room temperature \cite{indium};
 the ultimate shear stress for solder bumps is in
the 25-40 MPa range, while for In is 6.1 MPa. 

Excellent results on detector assemblies have been obtained on prototypes for the ATLAS and CMS
pixel detectors and have been reported at this conference
\cite{gemme,roland,polina}; the single bump failure has been measured
to be at ${\rm 10^{-5}}$ level and the system reliability is under investigation.

\subsection{Conclusion}
Flip chip technologies based on both solder and Indium bumps have been proven to be a solution 
for detector assembly. The statistical single bump failure rate is well within
the specifications. The choice between either of the two technologies will possibly be defined
by the long term reliability, the thermal compliancy of the assemblies and by the 
quality of the process control, i.e. by its repeatibility.

\section{Power lines and signal bus}
After the front-end chips are bump bonded, the assembly of a module is completed 
routing the power and digital lines,  
housing the controller chip and eventually temperature sensors and 
the opto-package for the interface to the far-end electronics.
This may be accomplished  by two different approaches:
\begin{itemize}
\item  the "flex" hybrid technology, where routing is defined in a 
multilayer polyimide film glued on the substrate.
The passive components are connected with standard surface mount technology and
wire bonds interconnect the traces to the I/O pads of the front-end chips. The controller chip
may be connected either by wire bonds or flip chip. This approach was used in DELPHI \cite{pierre}
and it is considered the baseline option for the pixel detectors at the LHC experiments and BTeV, 
reporting the latest developments at this conference \cite{cms_ge,pat,zimm}
\item the Multi Chip Module with Deposited Dielectric (MCM-D) technology, 
a robust and monolithic approach based on thin film technology \cite{izm1}.
In MCM-D, routing is again defined in a multilayer film but grown on the substrate,
where the dielectric is deposited by spin coating and the traces are electroplated.
Passive components may be integrated in thin film technology and the
chip I/O pads are interconnected by bump bonding. This approach is being developed for the
ATLAS pixel detector; the basic principles and the latest developments have also been
reported at this conference \cite{mcm}. 
\end{itemize}
The two solutions are similar as far as the passive material added to the system 
(${\rm \approx 0.1\%~X_{0}}$ \cite{tdr}) and opposite in any other respect. 
Flex hybrids are a mature industrial technology
and the requirements for the generic detector assembly fit the standards. Wire bonding
reliability is a concern but it is not expected to be critical. The main issue is the very large
mismatch between the coefficients of thermal expansion of flex (${\rm \approx 45 \times
10^{-6}~^{o}C^{-1}}$) and Silicon (${\rm 2.5 \times 10^{-6}~^{o}C^{-1}}$) that can induce 
a severe stress on the bumps moving from the assembly to the operating temperatures, i.e.
from ${\rm 20^{o}C}$ to ${\rm \approx -10^{o}C}$ \cite{polina}.
The MCM-D technology offers the advantages and disadvantages of a monolithic approach.
On one hand, no wire bonds are foreseen and the assembly scheme minimizes the stress 
on the bumps, as the multilayer is to be grown on the junction side of the sensor. 
On the other hand, the requirements on the single bump failure rates are tighter, as the 
front-end I/O pad are also bump bonded; moreover, each sensor cell is connected to 
the mating electronics cell by a staircase of feed through connections also requiring a 
high quality assurance. Last but not least, the MCM-D is necessarily tested after being grown
on the sensor substrate so the possibility to rework the substrate in case of failure is a 
must in terms of yield optimization.
The current results demonstrate the integrated solution is certainly appealing but the process
optimization is still under way, both in industrial applications and in detector assembly.

\section{Conclusions}
The next generation of pixel detector trackers represents a real challenge as every system
characteristics is at the edge of the current technology. Moreover, the total area is such that
the proposed solutions have also to guarantee a high production yield. Hybridization is not
less demanding: suitable technologies exist and quality assessment is on the way, with a full
qualification by the production starting soon.

\end{document}